\documentclass[12pt,a4paper]{article}
\usepackage{amsmath, amssymb}
\usepackage{amsfonts}
\usepackage[mathscr]{eucal}
\usepackage{ascmac}
\usepackage[dvips]{graphicx}
\usepackage{subfigure}
\usepackage[charter]{mathdesign}
\usepackage{bm}
\usepackage{cite}
\usepackage{here}
\usepackage{color}
\usepackage{comment}

\usepackage[height=25cm,width=16.5cm]{geometry}
\begin{document}
\renewcommand{\thefootnote}{$\clubsuit$\arabic{footnote}}
%%%%%%%%%%%%%%%%%%%%%%%%%%%%%%%%%%%%%%%%%%%
\def\a{\alpha}
\def\b{\beta}
\def\c{\varepsilon}
\def\d{\delta}
\def\e{\epsilon}
\def\f{\phi}
\def\g{\gamma}
\def\h{\theta}
\def\k{\kappa}
\def\l{\lambda}
\def\m{\mu}
\def\n{\nu}
\def\p{\psi}
\def\q{\partial}
\def\r{\rho}
\def\s{\sigma}
\def\t{\tau}
\def\u{\upsilon}
\def\v{\varphi}
\def\w{\omega}
\def\x{\xi}
\def\y{\eta}
\def\z{\zeta}
\def\D{\Delta}
\def\G{\Gamma}
\def\H{\Theta}
\def\L{\Lambda}
\def\F{\Phi}
\def\P{\Psi}
\def\S{\Sigma}

\def\der{\partial}
\def\o{\over}
\def\beq{\begin{eqnarray}}
\def\eeq{\end{eqnarray}}
\newcommand{\gsim}{ \mathop{}_{\textstyle \sim}^{\textstyle >} }
\newcommand{\lsim}{ \mathop{}_{\textstyle \sim}^{\textstyle <} }
\newcommand{\vev}[1]{ \left\langle {#1} \right\rangle }
\newcommand{\bra}[1]{ \langle {#1} | }
\newcommand{\ket}[1]{ | {#1} \rangle }
\newcommand{\EV}{ {\rm eV} }
\newcommand{\KEV}{ {\rm keV} }
\newcommand{\MEV}{ {\rm MeV} }
\newcommand{\GEV}{ {\rm GeV} }
\newcommand{\TEV}{ {\rm TeV} }
\def\diag{\mathop{\rm diag}\nolimits}
\def\Spin{\mathop{\rm Spin}}
\def\SO{\mathop{\rm SO}}
\def\O{\mathop{\rm O}}
\def\SU{\mathop{\rm SU}}
\def\U{\mathop{\rm U}}
\def\Sp{\mathop{\rm Sp}}
\def\SL{\mathop{\rm SL}}
\def\tr{\mathop{\rm tr}}
\def\Dterm{_{\theta^2\tilde{\theta^2}}}
\def\Fterm{_{\theta^2}}
\def\F*term{_{\tilde{\theta}^2}}
%%%%%%%%%%%%%%%%%% DEFINITION %%%%%%%%%%%%%%%%%%%
\def\mi{m_{\phi}}
\def\mpl{M_{\rm pl}}
%%%%%%%%%%%%%%%%%% DEFINITION %%%%%%%%%%%%%%%%%%%

\def\IJMP{Int.~J.~Mod.~Phys. }
\def\MPL{Mod.~Phys.~Lett. }
\def\NP{Nucl.~Phys. }
\def\PL{Phys.~Lett. }
\def\PR{Phys.~Rev. }
\def\PRL{Phys.~Rev.~Lett. }
\def\PTP{Prog.~Theor.~Phys. }
\def\ZP{Z.~Phys. }

%%%%%%%%%%%%%%%%%%%%%%%%%%%%%%%%%%%%%%%%%%%%%%%%%%%%%%%%%%%%%%%%%%%%

%%Eq. #
\makeatletter
\@addtoreset{equation}{section}
\def\theequation{\thesection.\arabic{equation}}
\makeatother
\newcommand{\Slash}[1]{{\ooalign{\hfil/\hfil\crcr$#1$}}} 

%%%%% the following command must be removed when submitting.
\newcommand{\TODO}[1]{{$[[ \clubsuit\clubsuit$ \bf #1 $\clubsuit\clubsuit ]]$}}
\newcommand{\kmrem}[1]{{\color{red} \bf $[[ $ KM: #1$ ]]$}}
\newcommand{\knrem}[1]{{\color{blue} \bf $[[ $ KN: #1$ ]]$}}
\newcommand{\mtrem}[1]{{\color{green} \bf $[[ $ MT: #1$ ]]$}}
%%%%%%%%%%%%%%%%%%%%%%%%%%%%%

\begin{titlepage}

\begin{flushright}
%\large{\textit{Internal Note}}
UT-14-22
\end{flushright}

\vskip 4cm

\begin{center}
{\huge \bfseries
Correspondence of I- and Q-balls\\[.7em]
as
Non-relativistic Condensates
}

\vskip .9in
{\large
Kyohei Mukaida
and Masahiro Takimoto
}

\vskip .35in
\begin{tabular}{ll}
& \!\! {\em Department of Physics, Faculty of Science, }\\
& {\em University of Tokyo,  Bunkyo-ku, Tokyo 133-0033, Japan}\\[.5em]
\end{tabular}

\vskip .95in

\begin{abstract}
%In the non-relativistic limit,
%\mtrem{With a non-relativistic configuration,}
If a real scalar field is dominated by non-relativistic modes, then
%a real scalar field 
it approximately conserves its particle number 
and 
obeys an equation 
that 
governs
a complex scalar field theory with a conserved global U(1) symmetry.
% in the non-relativistic limit.
From this fact, it is shown that the I-ball (oscillon) can be naturally understood as a projection (\textit{e.g.}, real part) 
of the non-relativistic Q-ball solution.
In particular, we clarify that the stability of the I-ball is guaranteed by the U(1) symmetry
in the corresponding complex scalar field theory
as long as the non-relativistic condition holds.
We also discuss the longevity of I-ball from the perspective of the complex scalar field
in terms of U(1) charge violating processes.
\end{abstract}

\end{center}

\end{titlepage}

\newpage

%\tableofcontents

%%%%%%%%%%%%%%%%%%%%%%%%%%%%%%%%%%%%%%%%%%%%%%%%%%
\section{Introduction}
\label{sec:}
%%%%%%%%%%%%%%%%%%%%%%%%%%%%%%%%%%%%%%%%%%%%%%%%%%

Condensates of scalar fields play important roles in the early stage of the Universe.
One of the prominent examples is inflation\cite{Guth:1980zm,Linde:1981mu},
which not only solves the horizon/flatness problems and dilutes unwanted relics by
the accelerated expansion of Universe,
but also explains seeds of primordial density fluctuations, observed by Ref.~\cite{Ade:2013zuv}.
The quasi-exponential expansion can be driven by the vacuum expectation value of some scalar field(s)
which is so-called inflaton.

Scalar fields other than the inflaton may also acquire large expectation values during inflation.
For instance, if the scalar field is charged under the baryonic U(1) symmetry, 
it can produce the baryon asymmetry of the Universe
via the Affleck-Dine (AD) mechanism~\cite{Affleck:1984fy}.
The baryon asymmetry is produced at the beginning of the coherent oscillation of the AD field via
the enhanced baryon number breaking term due to its large expectation value.

Some of these coherently oscillating scalar fields eventually fall into long-lived localized objects,
if they are associated with conserved quantities: topological/non-topological charge.
Once these solitons are formed, they may significantly affect the evolution of the Universe due to their longevity.
For instance,
it is known that the coherently oscillating AD field fragments into spherically localized lumps,
the so-called Q-balls,
after the growth of inhomogeneous fluctuations due to the instability of the coherent oscillation~\cite{Coleman:1985ki} .
And the formation of Q-balls alters the AD baryogenesis scenario crucially~
\cite{Kusenko:1997si,Enqvist:1997si,Enqvist:1998en,Kasuya:1999wu,Kasuya:2000wx,Kasuya:2001hg}.
The stability of these solitons is guaranteed by the conserved topological/non-topological charge.
In the case of the Q-balls, it is stable owing to the conserved U(1) symmetry.

The simplest example of scalar field theory may be 
a single real scalar field whose potential is dominated by a quadratic term.
Despite its simplicity, it can be one of the candidates of inflaton that is favored 
by the recent observation~\cite{Ade:2014xna}.
Apparently, such a single real scalar field cannot form stable lumps because there are no evident conserved charges.
However, contrary to the naive expectation,
many numerical studies reveal that it can actually fragment into long-lived spherical objects,
the so-called oscillon~\cite{Bogolyubsky:1976nx,Gleiser:1993pt,Copeland:1995fq,Gleiser:1999tj,
Honda:2001xg,Gleiser:2004an,Fodor:2006zs,Hindmarsh:2006ur,Saffin:2006yk,
Fodor:2008es,
Fodor:2008du,Gleiser:2009ys,Amin:2010jq,
Amin:2011hj,Salmi:2012ta,Andersen:2012wg,Kawasaki:2013hka,
Saffin:2014yka},
if the potential is nearly but shallower than quadratic.\footnote{
{In this paper, we concentrate on the simple case of oscillon whose effective potential
is dominated by the quadratic term.}
}
In Ref.~\cite{Kasuya:2002zs} ,
it was shown that the stability of oscillons is guaranteed by the conservation of
the adiabatic charge $I$
which is approximately conserved if the potential is nearly quadratic.
Since the adiabatic charge, $I$, plays the same role as the U(1) charge, $Q$, in the case of the Q-balls, 
they are dubbed as I-balls.

In this paper, we revisit the stability of I-balls by reinterpreting the conservation of the adiabatic charge.
Since the adiabatic charge is approximately conserved for nearly quadratic potentials,
the adiabatic invariance implies that I-balls can be well described 
in terms of the non-relativistic real scalar field theory.
We call the field configurations such that
the dominant part of scalar field is non-relativistic 
and other relativistic part is too small to affect the dynamics of the system,
as ``non-relativistic configuration''.
Interestingly, %in the non-relativistic \mtrem{configuration},
the non-relativistic real scalar field approximately conserves its number and
obeys the equation
which governs
a complex scalar field with a conserved U(1) symmetry
%in the non-relativistic \mtrem{configuration}
~\cite{Berges:2014xea,Davidson:2014hfa}.
Therefore,
from the perspective of this corresponding complex scalar field theory,
the stability of I-balls is guaranteed by the U(1) symmetry 
associated with the number conservation in the original real scalar one
in the non-relativistic configuration.
In this sense,
I-balls are stabilized by the symmetry,
which is not clear in the language of the original real scalar field theory.
As a result,
the I-ball solution can be understood as the projection (\textit{e.g.,} real part) of the Q-ball one.
We also discuss the duration time while the non-relativistic %\mtrem{configuration} 
condition holds after the I-ball formation
from the viewpoint of corresponding complex scalar field,
and find that it is much longer than the oscillation time scale of the scalar field.

In Sec.~\ref{sec:Non}, we will see in the non-relativistic configuration, 
a real scalar field theory can be embedded into a complex one with
a conserved U(1) symmetry.
In Sec.~\ref{sec:I/Q}, we show 
that the energetically favored solution of the non-relativistic 
real scalar field can be described by the projection of Q-ball and see
the correspondence of I-ball and Q-ball solution.
In terms of charge violating processes within the corresponding
complex scalar theory,
we also discuss the robustness of the non-relativistic configuration %condition
after the formation of I-ball
and estimate the lifetime of I-ball in Sec.~\ref{sec:stb}.
Sec.~\ref{sec:conc} is devoted to the conclusion.

%%%%%%%%%%%%%%%%%%%%%%%%%%%%%%%%%%%%%%%%%%%%%%%%%%
\section{Non-relativistic condition and number conservation}
\label{sec:Non}
%%%%%%%%%%%%%%%%%%%%%%%%%%%%%%%%%%%%%%%%%%%%%%%%%%

In this section, we show that
the real scalar field theory can be embedded into the complex one.
In particular, there is an approximate $\text{U}(1)$ symmetry 
for the non-relativistic configuration of the real scalar field.
In order to discuss the non-relativistic configuration,
let us first clarify the setup which we study throughout this paper.
Consider a generic real scalar field theory of $\phi$ with an interaction term $V (\phi)$.
The Lagrangian density is then given by\footnote{
In this paper, we regard the Lagrangian as an effective one.
Thus, the potential and also the mass term depend on the setup such as
the background temperature, the amplitude of the field, for instance.
We assume that the thermal dissipation effects on the evolution of  $\phi$
are negligible for simplicity, otherwise one has to take them into account.
}
\begin{align}
	\mathcal{L} = \frac{1}{2} \der_\mu \phi \der^\mu \phi - \frac{1}{2} m^2 \phi^2 - V (\phi).
\end{align}
Since we are interested in the regime where the quadratic term dominates the potential,
here we extract the mass term explicitly. 
The equation of motion following from this action reads
\begin{align}
\label{eq:eom_r}
	0 = \left( \Box + m^2 \right) \phi + V' (\phi).
\end{align}

To clarify the consequence of approximate number conservation in the non-relativistic configuration,
it is convenient to recast the equation in terms of the complex scalar field theory. 
By using a complex scalar $\Phi$, one can express Eq.~\eqref{eq:eom_r} as
\begin{align}
	0 &= \left( \Box + m^2 \right) \Re[\Phi] + V' \left(\Re [\Phi]\right) 
	= \Re \left[ \left( \Box + m^2 \right) \Phi +  \frac{\der V_\text{U(1)} (|\Phi|)}{\der \Phi^\dag}  
	+ \frac{\der V_\text{B} (\Phi,\Phi^\dag)}{\der \Phi^\dag} \right],
\end{align}
where $\phi = \Re[\Phi]$ and $V_\text{U(1)/B}$ is a real function.
In appendix \ref{sec:CC}, we 
concretely show how to
construct $V_\text{U(1)/B}(\Phi)$ from $V(\phi)$. 
The solution of Eq.~\eqref{eq:eom_r} can be obtained from
the real part of the solution of
\begin{align}
\label{eq:eom_c}
	0 &= \left( \Box + m^2 \right) \Phi + \left[ \frac{\der V_\text{U(1)} (|\Phi|)}{\der \Phi^\dag}  
	+ \frac{\der V_\text{B} (\Phi,\Phi^\dag)}{\der \Phi^\dag}  \right],
\end{align}
as $\phi = \Re [\Phi]$
with an appropriate initial condition.
This equation of motion can be obtained from the following action of a complex scalar field theory:
\begin{align}
	\mathcal L = \der_\mu \Phi^\dag \der^\mu \Phi - m^2 |\Phi|^2 - V_\text{U(1)} (|\Phi|) - V_\text{B} (\Phi, \Phi^\dag).
\end{align}
Importantly, the force in Eq.~\eqref{eq:eom_c} can be divided into two terms;
(i) global U(1) conserving term $\der V_\text{U(1)}(|\Phi|) / \der \Phi^\dag$ 
and (ii) global U(1) breaking term $\der V_\text{B}(\Phi, \Phi^\dag) / \der \Phi^\dag$,
where the global U(1) transformation is defined as $\Phi \mapsto e^{i \theta} \Phi$.
Therefore, if one can neglect the second term for some reasons,
the equation of motion respects the global U(1) symmetry.
As we will see, the non-relativistic configuration implies the negligence of the second term
that describes charge violating processes.

To be concrete, it is instructive to consider a polynomial potential:
\begin{align}
\label{eq:phi4}
	V (\phi) = - \frac{\lambda}{4} \phi^4 + \frac{g}{6m^2}\phi^6,
\end{align}
where $\lambda,g>0$ and we assume that there is no metastable vacuum.
In this case, the U(1) conserving/breaking potential of the corresponding complex scalar field theory
can be obtained as 
\begin{align}
	V_\text{U(1)} (|\Phi|) &= - \frac{3 \lambda}{8} |\Phi|^4+\frac{5g}{24m^2}|\Phi|^6,\\[10pt]
	V_\text{B} (\Phi, \Phi^\dag) &= - \frac{\lambda}{16} \left( \Phi^4  + {\Phi^\dag}^4 \right)
	+\frac{g|\Phi|^2}{16m^2}\left(\Phi^4+{\Phi^\dag}^4\right)
	.
\end{align}
As one can see, $V_\text{U (1)}$ respects the U(1) symmetry, $\Phi \to e^{i \theta} \Phi$, but
$V_\text{B}$ breaks it explicitly.

Now we are in a position to consider the non-relativistic configuration.
We call the field configuration as the non-relativistic one if the following condition holds.
The main part of oscillation energy is dominated solely by the mass term, $E = m + p^2/(2m) + \cdots$,
and hence 
we can factor out this oscillation and separate the other small part as 
\begin{align}
\label{eq:ansatz}
	\Phi = e^{-i m t} \Psi+\delta \Phi,
\end{align}
where $\delta \Phi$ contains 
rapidly oscillating parts but
is assumed to be so small that it cannot affect the dynamics of the dominant part $\Psi$:
$|\Psi| \gg |\delta\Phi|$.
This condition holds if the momentum 
of $\Psi$ is small $p < m$, the potential is dominated by the mass term
$|V'_\text{U(1)} (\Psi_0) |\ll m^2 \Psi_0$ with $\Psi_0$ being a typical amplitude 
and effects of $\delta \Phi$ on the dynamics of $\Psi$ are negligible for a sufficient long time.
Here $\Psi$ is a slowly varying field: 
\begin{align}
	\left| \Psi \right| \gg \left| \frac{\der \Psi}{m \der t} \right| \gg \left| \frac{\der^2 \Psi}{m^2 \der t^2} \right|.
\end{align}
This condition implies that the initial condition of $\Psi$ should be related with $\phi$ as
\begin{align}
	\left. \Re\Psi \right|_\text{ini} \simeq \left.\phi \right|_\text{ini};~~ 
	m \left. \Im \Psi  \right|_\text{ini} \simeq \left.\dot \phi \right|_\text{ini}.
\end{align}
Note that the slowly varying field $\Psi$ can evolve with a time scale smaller than 
the mass term $m$ as explicitly introduced later in Sec.~\ref{sec:I/Q}.\footnote{
More precisely, as we will see later,
the oscillation time scale of the dominant mode becomes smaller than the mass term
for energetically favored configurations:
$m - \mu < m$ with $\Psi = e^{i\mu t} \tilde \Psi$.
}
%\mtrem{
%The potential is dominated by the mass term if the following condition holds
%%
%\begin{align}
%	\left |V'_\text{U(1)} (\Phi_0)
%	\text{ or } V'_{\rm B} (\Phi_0)
%	\right| \ll m^2 \Phi_0,
%\end{align}
%%
%with $\Phi_0$ being a typical amplitude and
%which we call as non-relativistic condition.
%}
%In addition $\delta \Phi$ must be small for sufficiently long time
%though there is no need for it to be non-relativistic.
 
Inserting the ansatz Eq.~\eqref{eq:ansatz} and neglecting higher derivatives of the slowly varying field,
one finds
\begin{align}
\label{eq:ggp}
	0 = e^{-i mt} \left[ - 2 i m \frac{\der}{\der t} - \nabla^2 + U_\text{U(1)} (|\Psi|) \right] \Psi;~~~
	U_\text{U(1)} (|\Phi|) \equiv \frac{1}{2 |\Phi|} V_\text{U(1)}' (|\Phi|).
\end{align}
It is noticeable that
the U(1)  breaking term $\der V_\text{B} (e^{-imt} \Psi, e^{imt} \Psi^\dag) / \der \Phi^\dag$ contains
more rapidly oscillating term than $e^{-imt}$, and hence
this charge violating term averages to zero
as long as the above non-relativistic condition holds.
More precisely, for instance, in the case of polynomial potential, 
the U(1) violating term contains $\der V_\text{B} / \der \Phi^\dag \supset - e^{3imt} (\lambda/4) {\Psi^\dag}^3$, which cannot be compensated by a slowly varying field $\Psi$,
rather by the small rapidly oscillating term $\delta \Phi$ [See Eq.~\eqref{eq:cmp} in Sec.~\ref{sec:stb}].
Roughly speaking, as long as the non-relativistic condition holds,
charge violating processes are both negligible.
We will return to this issue after the formation of I-ball in Sec.~\ref{sec:stb}.

To sum up,
in the non-relativistic condition, 
the charge violating term $V_\text{B}$ can be neglected, and as a result,
the solution of the real scalar field can be obtained from Eq.~\eqref{eq:ggp},
which describes the dynamics of 
a non-relativistic complex scalar field theory with a U(1) symmetry.
The Lagrangian density of this complex scalar field theory reads
\begin{align}
\label{eq:complex_th}
	\mathcal{L} = \der_\mu \Phi^\dag \der^\mu \Phi - m^2 |\Phi |^2 - V_\text{U(1)} (|\Phi|).
\end{align}
One can clearly see that the non-relativistic condition
implies the restoration of U(1) symmetry
associated with the approximate number conservation of the original real scalar field theory.
Therefore, a quasi-stable non-topological soliton can be formed even in the real scalar field theory,
as we see in the following section.
In the next section,
we show that a stationary localized solution of Eq.~\eqref{eq:ggp} is nothing but the Q-ball
and its real part is the I-ball
 \footnote{{Strictly speaking, the obtained solution of Eq.~\eqref{eq:ggp}
 does not exactly satisfy the original equation
 of motion Eq.~\eqref{eq:eom_r}.
 In general, there exist fast oscillating corrections $\delta \Phi$ which is found to be small
 in Sec.\ref{sec:stb} a posteriori.
 Thus, what one can obtain from Eq.~\eqref{eq:ggp} is the dominant part of the I-ball solution.
}}.

%%%%%%%%%%%%%%%%%%%%%%%%%%%%%%%%%%%%%%%%%%%%%%%%%%
\section{I- and Q-balls}
\label{sec:I/Q}
%%%%%%%%%%%%%%%%%%%%%%%%%%%%%%%%%%%%%%%%%%%%%%%%%%
In this section,
we study energetically favored non-relativistic configurations.
%when the non-relativistic condition holds.
Let us seek a non-trivial stationary localized solution of
\begin{align}
\label{eq:ggp2}
	0 = \left[ - 2 i m \frac{\der}{\der t} - \nabla^2 + U_\text{U(1)} (|\Psi|) \right] \Psi;~~~
	U_\text{U(1)} (|\Psi|) \equiv \frac{1}{2 |\Psi|} V_\text{U(1)}' (|\Psi|).
\end{align}
Soon,
we will identify the solution as the Q-ball with a potential dominated by a quadratic term.
As a result, one can simultaneously obtain the non-relativistic solution of Eq.~\eqref{eq:eom_r} as $\phi=\Re[e^{-imt}\Psi]$
that is turned out to be the I-ball in the following.

Since a spherically asymmetric solution costs energy~\cite{Coleman:1977th},
$\Psi$ depends on a radius $r$ rather than $\bm{x}$.
Therefore all one has to do is to find the bounce solution:
\begin{align}
\label{eq:bounce}
	\left[ \frac{\der^2}{\der r^2} + \frac{2}{r} \frac{\der}{\der r} \right] \tilde\Psi(r)
	= - V_\mu' (\tilde\Psi);~~
	V_\mu (\tilde \Psi) \equiv \left[ -  \mu m \tilde \Psi^2 - \frac{1}{2}V_\text{U(1)}(\tilde\Psi) \right],
\end{align}
with boundary conditions
\begin{align}
\label{eq:bdry}
	\lim_{r \to 0} \tilde \Psi' = 0;~~ \lim_{r \to \infty} \tilde \Psi = 0.
\end{align}
Here we rewrite $\Psi (t,r) = e^{i\mu t + \theta} \tilde\Psi (r)$ with $|\mu| \ll m$.
If $m-\mu$ is smaller than the frequency outside the localized solution, the solution is
energetically favored.
As is often the case,
Eq.~\eqref{eq:bounce} can be regarded as a one dimensional equation of motion; with
a time $r$, a friction term $2/r$, and a potential 
$V_\mu (\tilde \Psi)$.
Hence, the bounce solution exists if the following condition is met:
\begin{align}
\label{eq:cond}
	m^2 + \min \left[ \frac{V_\text{U(1)} (\tilde \Psi)}{\tilde \Psi^2} \right] 
	< \left( m^2 - 2 \mu m \right) < \left[ m^2 + \frac{1}{2}  V_\text{U(1)}'' (0) \right],
\end{align}
which is the same as the ordinary condition for the existence of Q-ball solution 
because $\omega^2 \equiv (m-\mu)^2 \simeq m^2 - 2 \mu m$.
Note that owing to the potential shallower than the quadratic one,
a localized solution can exist.
A balance between the pressure and the attractive force
leads to the following relation for a typical wall width, $L$, and an amplitude of Q-ball, $\Phi_0$, as
\begin{align}
\label{relat}
	L^2 V_\text{U(1)}' (\Phi_0) / \Phi_0 \sim 1.
\end{align}
Together with condition that the mass term dominates the potential, $|V'_\text{U(1)} (\Phi_0) |\ll m^2 \Phi_0$,
Eq.~\eqref{relat} implies
\begin{align}
\label{eq:small}
	1/\epsilon&\equiv L\omega \simeq Lm \gg 1,
\end{align}
which is consistent with $p \ll m$.

Now it is obvious that the non-trivial stationary localized solution of Eq.~\eqref{eq:ggp2}
is the so-called Q-ball solution of Eq.~\eqref{eq:complex_th}.
This is because the governing equation Eq.~\eqref{eq:bounce}, 
the boundary conditions Eq.~\eqref{eq:bdry} and the condition Eq.~\eqref{eq:cond}
are the same as those of the Q-ball.
The corresponding solution for the real scalar field is obtained from its real part.
It is so-called I-ball since Eqs.~\eqref{eq:bounce},
\eqref{eq:bdry} and \eqref{eq:cond}
are essentially the same as those given in Ref.~\cite{Kasuya:2002zs}.

To make our discussion concrete, 
we consider
a polynomial potential and
a gravity mediation type potential 
in the following.
First, let us move back to the case of polynomial potential [Eq.~\eqref{eq:phi4}]:
\begin{align}
	V (\phi) = - \frac{\lambda}{4} \phi^4+\frac{g}{6m^2}\phi^6.
\end{align}
As explained previously,
the potential of the corresponding complex scalar field theory with a U(1) symmetry
is given by  
\begin{align}
	V_\text{U(1)} (|\Phi|) = - \frac{3 \lambda}{8} |\Phi|^4+\frac{5g}{24m^2}|\Phi|^6.
\end{align}  
Hence, Eq.~\eqref{eq:bounce} can be expressed as
\begin{align}
	\left[ \frac{\der^2}{\der r^2} + \frac{2}{r} \frac{\der}{\der r} \right] \tilde \Psi = 
	2 \mu m \tilde \Psi - \frac{3 \lambda}{4} \tilde \Psi^3+
	\frac{5 g}{8m^2} \tilde \Psi^5.
\end{align}
In this case, $\tilde{\Psi}(r=0)$ has an upper limit 
$\Phi_{\rm cr}\equiv \sqrt{9\lambda/(5g)}\,m$. In order to see its typical behavior,
we numerically solve the bounce equation.

\begin{figure}[th]
\centering
\subfigure[{\bf Left Panel}]{
\includegraphics[width=0.46\columnwidth,clip]{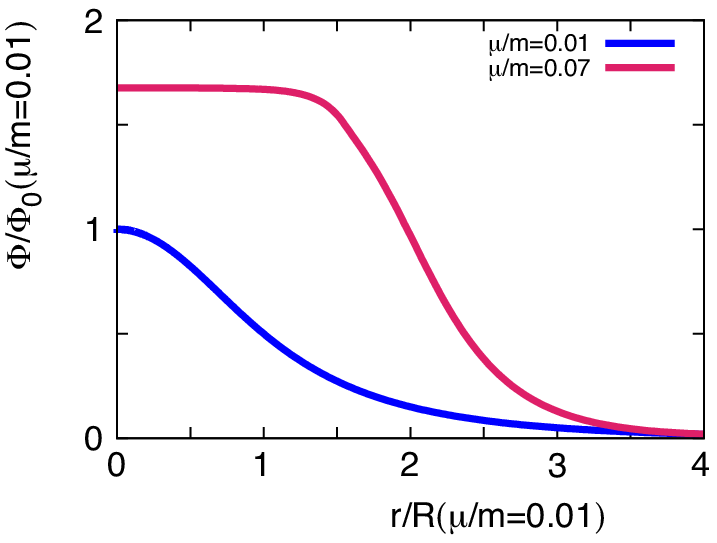}
  \label{fig:prof}
}
\subfigure[{\bf Right Panel}]{
   \includegraphics[width=0.48\columnwidth,clip]{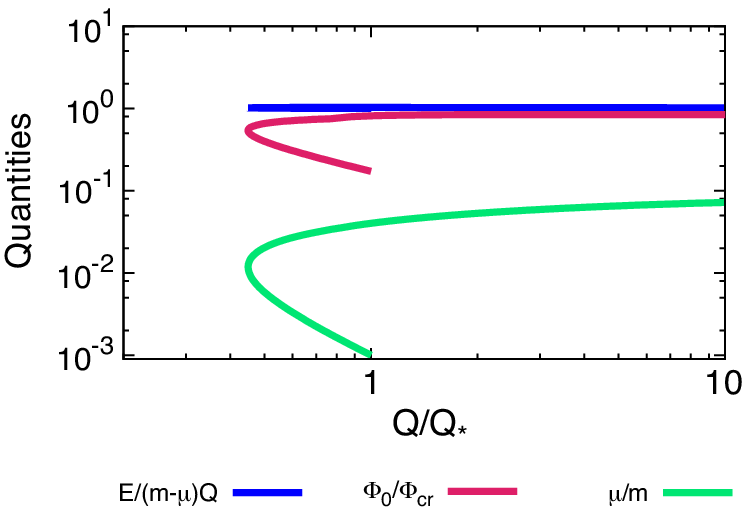}
   \label{fig:fix}
 }
 \caption{\small {\bf  Left Panel} shows the profile of I/Q-ball for $\mu=0.01 m$ and $\mu=0.07m$
 for $\lambda=0.1,g=\lambda^2$ normalized by values at $\mu=0.01m$.
 {\bf Right Panel} shows several dimensionless quantities of I/Q-ball for $\lambda=0.1,g=\lambda^2$ as a function of normalized 
   charge $Q/Q_\ast$}
\end{figure}

We set parameters as $\lambda=0.1,g=\lambda^2$.\footnote{
We have numerically checked the wall width is large enough compared to the mass inverse
for these parameters. }
Fig.~\ref{fig:prof} shows a profile of I/Q-ball for $\mu=0.01m$ and $\mu=0.07m$ normalized
by the values at $\mu=0.01m$
where we define a typical size of I/Q ball $R$ at the point $\Phi(R)\equiv\Phi_0/2$
with $\Phi_0$ being the field value at $r=0$. 
Fig.~{\ref{fig:fix}} shows several dimensionless quantities
as a function of $Q/Q_{\ast}$ where we denote the charge of corresponding Q-ball
for $\mu=10^{-3}m$ as $Q_\ast$.
From blue line, one can see that the relation $E\simeq \omega Q \equiv (m-\mu) Q$
holds with a good accuracy.
$\Phi_0/\Phi_{\rm cr}$ shows the amplitude of the Q/I-ball normalized by
$\Phi_{\rm cr}$.
The green line indicates $\mu$ normalized by $m$.
The dependence of $\omega\equiv m-\mu$ on $Q$ is important for the 
stability of the I-ball as is discussed in Sec.~\ref{sec:stb}. 
One can see that for relatively small $\mu$ (the lower branch), $Q$ drops down but
$\Phi_0$ increases as the $\mu$ increases.
On the other hand, for relatively large $\mu$ (the upper branch),  as 
the $\mu$ increases; $Q$ increases,
$\Phi_0$ becomes close to $\Phi_{\rm cr}$ and the profile becomes flatter. 
We will call the former regime as ``narrow regime'' and
the latter one as ``flat-top regime'' as in \cite{Amin:2010jq}.
In other words, the former/latter solution is nothing but so-called
thick/thin wall regime of the bounce.
These results are consistent with \cite{Amin:2010jq,Enqvist:2003zb}.
Roughly speaking, in the narrow regime attractive force due to the quartic potential
balance with the pressure,
but in the flat-top regime repulsive force due to the sixtic potential
stops the growth of $\Phi_0$ at the center.

Next, suppose that the real scalar field theory has {the} following potential:
\begin{align}
	V (\phi) = - \frac{1}{2} m^2 (\phi_0) \phi^2K \ln \left[ \frac{\phi^2}{\phi_0^2} \right],
\end{align}
where $K$ comes from a one-loop correction which is assumed to be small and positive,
$0 < K \ll 1$, and $m^2 (\phi_0)$ is the mass defined at the scale $\phi_0$.
Here and hereafter we write the mass term as
$m^2 = m^2 (\phi_0)$ unless 
otherwise stated.
The corresponding complex scalar field theory with a conserved global U(1) symmetry has 
a following potential:
\begin{align}
	V_\text{U(1)} (|\Phi|) = - m^2 |\Phi|^2 K \ln \left[ \frac{|\Phi|^2}{\Phi_0^2} \right]; ~~
	\Phi_0^2 \equiv c \phi_0^2,
\end{align}
with $c$ being a numerical constant given by $4/e$.
Thus, Eq.~\eqref{eq:bounce} reads 
\begin{align}
\label{eq:bounce_grv}
	\left[ \frac{\der^2}{\der r^2} + \frac{2}{r} \frac{\der}{\der r} \right] \tilde \Psi =
	- \left( m^2 - 2 \mu m \right) \tilde \Psi + m^2 \tilde \Psi \left( 1 - K - K \ln \left[ \frac{\tilde \Psi^2}{\Phi_0^2} \right] \right).
\end{align}
Since this potential satisfies the condition Eq.~\eqref{eq:cond} as long as $K > 0$,
there is a non-trivial solution with boundary conditions Eq.~\eqref{eq:bdry}.
It is known that the solution can be well approximated by the Gaussian form in this case.
Thus, the Q-ball solution has the following form
\begin{align}
	\Phi_Q(t,r) = e^{-i(m-\mu)t}\tilde \Psi_G (r);~~ \tilde \Psi_G (r) = \Phi_0 \exp \left[ - \frac{r^2}{2 R^2} \right],
\end{align}
where
\begin{align}
\label{eq:sol_grav}
	mR = K^{- 1/2}; ~~ \mu = - K m,
\end{align}
which satisfy the non-relativistic condition:
$1/R \ll m$ and $\mu \ll m$.
Note that this relation is consistent with the rough estimation given in Eq.~\eqref{relat}.
One can see that if the amplitude is large enough,
the Q-ball solution is energetically favored 
by considering 
the effective mass outside the Q-ball solution.
Suppose that the typical energy scale outside the Q-ball solution is $\Lambda$,
the effective mass outside $m(\Lambda)$ can be written as
\begin{align}
	m^2(\Lambda)\simeq m^2(\Phi_0)\left(1+K\ln \left[
	\frac{\Phi_0^2}{\Lambda^2}
	\right]\right).
\end{align}
If the amplitude $\Phi_0$ is larger than $\Lambda$ so that $\ln (\Phi_0/\Lambda)>1$, 
the condition $m(\Phi_0)-\mu<m(\Lambda)$ is satisfied
and the Q-ball solution is energetically favored for a positive $K$.
Contrary to the narrow regime in the polynomial potential,
since the radius is given by $R \propto m(\Phi_0)^{-1}$,
the frequency of  I/Q-ball, $\omega \sim m(\Phi_0)$, decreases as the charge increases,
$Q \propto m(\Phi_0) \Phi_0^2 R^3 \propto (\Phi_0/m(\Phi_0))^2$.
The I-ball solution reads
\begin{align}
	\phi_I (t,r) = \cos \left[ \left( m - \mu \right) t \right] \tilde \Phi_G (r).
\end{align}
Note that if one defines the amplitude of I-ball solution as a square root of oscillation time average
of $\phi_I^2$,
the amplitude becomes
\begin{align}
	\sqrt{\overline{\phi_I^2}} = \frac{\Phi_0}{\sqrt{2}} \exp \left[ - \frac{r^2}{2R^2} \right],
\end{align}
which is used in Ref.~\cite{Kasuya:2002zs}.

%%%%%%%%%%%%%%%%%%%%%%%%%%%%%%%%%%%%%%%%%%%%%%%%%%
\section{Stability of I-balls}
\label{sec:stb}
%%%%%%%%%%%%%%%%%%%%%%%%%%%%%%%%%%%%%%%%%%%%%%%%%%
Up to here, we have seen that
the I-ball can be regarded as a non-topological soliton associated with 
the approximate U(1) conservation as long as the non-relativistic 
condition is maintained.
However, after the formation of I-ball, if small fluctuations around the I-ball solution grow rapidly,
such relativistic modes break the condition and hence the stability of I-ball is lost.
Therefore, in this section, we discuss the robustness of the non-relativistic condition
around the I-ball solution,
which is verified if the effects from $V_{\rm B}(\Phi,\Phi^\dagger)$ term are negligible.

Before going into details,
it is instructive to recall the stability of Q-balls.
Once one finds the bounce solution $\tilde \Psi_B$, its energy in the complex scalar field theory
can be obtained from Eq.~\eqref{eq:complex_th} as
\begin{align}
\label{eq:e_q}
	E_\text{Q-ball} = \int 4\pi  r^2 dr  \left[ \left( m - \mu \right)^2 \tilde \Psi_B^2
	+ \left( \frac{\der}{\der r} \tilde \Psi_B \right)^2 + V_\text{U(1)} (\tilde \Psi_B) \right],
\end{align}
and its charge as
\begin{align}
\label{eq:c_q}
	Q = 2 (m - \mu) \int 4 \pi r^2 dr \tilde \Psi_B^2.
\end{align}
The Q-ball solution is stable only if its energy is smaller than that of free particles:
\begin{align}
\label{eq:stb_q}
	E_\text{Q-ball} < mQ.
\end{align}
Intuitively, this condition indicates that
the effective mass of the Q-ball solution is smaller than that of a free particle.
The shallowness of the potential ensures this condition.
For instance, in the case of the gravity mediation type potential,
its energy and charge are given by
\begin{align}
	E_\text{Q-ball} &= 2 \pi^{3/2} \left( 1 + \frac{5}{2}K \right) m^2 (\phi_0) \Phi_0^2, \\[10pt]
	Q &= 2 \pi^{3/2} \left( 1 + K \right) m (\phi_0) \Phi_0^2 R^3.
\end{align}
As explained in the previous section,
as long as the typical scale of background plasma $\Lambda$ is much smaller than $\phi_0$,
the condition \eqref{eq:stb_q} is satisfied
\begin{align}
	E_\text{Q-ball} \simeq m(\phi_0) Q \left( 1 + \frac{3}{2}K \right) < m(\Lambda) Q,
\end{align}
because $m(\Lambda) < m(\phi_0)$ for a positive $K$.

Then, let us discuss the stability of the corresponding I-ball.
Once it is guaranteed that the non-relativistic condition holds,
the I-ball solution can be obtained from that of the corresponding Q-ball
and hence it becomes (quasi-)stable as long as Eq.~\eqref{eq:stb_q} is met.
To justify this statement,  
we have to verify that small fluctuations around the I-ball solution, $\delta \Phi$,
does not grow violently due to the charge violating term $V_{\rm B}$,
since otherwise such modes immediately break the non-relativistic condition.
One can obtain the equation of motion for $\delta \Phi\equiv \Phi-\Phi_Q$
from Eq.~\eqref{eq:eom_c} as
\begin{align}
\label{eq:ful}
	\left(\Box+m^2\right)\delta\Phi=
	-\left. \frac{\partial V^\text{(int)}}{\partial \Phi^\dagger}\right|_{\Phi=\Phi_Q+\delta\Phi}
	+\left. \frac{\partial V_{\rm U(1)}}{\partial \Phi^\dagger}\right|_{\Phi=\Phi_Q},
\end{align}
where we denote the full interaction except for the mass term as
$V^\text{(int)}=V_{\rm U(1)}(|\Phi|)+V_{\rm B}(\Phi,\Phi^\dagger)$.

To make our discussion concrete,
let us for example see $\phi^4$ theory:
$V^\text{(int)} = - (3 \lambda / 8) |\Phi|^4 - (\lambda / 16) (\Phi^4+{\Phi^\dagger}^4)$. 
Then, the equation of motion for $\delta\Phi$ can be written as
\begin{align}
        \left(\Box+m^2\right)\delta\Phi = 
         \frac{3\lambda}{2}  |\Phi_Q|^2\delta \Phi
        + \frac{\lambda}{4} {\Phi^\dagger_Q}^3
        + \frac{3 \lambda}{4}
        \left(\Phi_Q^2+{\Phi_Q^\dagger}^2\right)\delta\Phi^\dagger+O(\delta\Phi^2),
\end{align}
at the leading order in $\delta \Phi$.
Note that $\delta\Phi=0$ does not satisfy the equation of motion
because the rapidly oscillating $V_B$ term, $\lambda {\Phi_Q^\dag}^3$, remains.
Thus, we separate $\delta \Phi$ into $\delta \Phi_{\rm cmp}$ and $\delta\Phi_{\rm flc}$, and
assume $\delta \Phi_{\rm cmp}$ compensates this term inside the I-ball.
In the case of the $\phi^4$ potential, $\delta \Phi_{\rm cmp}$ can be estimated as a solution of 
the following equation
\begin{align}
\label{eq:cmp}
          \left(\Box+m^2 - \frac{3 \lambda}{2} |\Phi_Q|^2 \right)\delta\Phi_\text{cmp} \simeq
          \frac{\lambda}{4} {\Phi^\dagger_Q}^3.
\end{align}
$\delta \Phi_\text{cmp}$ simply oscillates with a frequency $3 (m - \mu)$
inside the I-ball,
and does not grow violently.
A typical amplitude of the solution can be estimated as 
$|\delta \Phi_\text{cmp} | \sim (\lambda \Phi_0^2 / m^2 ) \Phi_0 \ll \Phi _0$,
which justifies the approximation we used, $|\Phi_Q| \gg |\delta \Phi|$.
Note that in general $|\delta \Phi_{\rm cmp}|\sim |V'_{\rm B}(\Phi_Q,\Phi^\dag_Q)|/m^2
\ll \Phi_0$ is ensured by the non-relativistic condition:
\begin{align}
	\left|V'_{\rm B}(\Phi_Q,\Phi^\dag_Q)\right|\sim
	\left|V'_{\rm U(1)}(|\Phi_Q|)\right|
	\ll m^2|\Phi_Q|.
\end{align}
Hence, $\Phi_\text{cmp}$ does not threaten the non-relativistic condition.
Here we omit effects from $\lambda (\Phi_Q^2 + {\Phi_Q^\dag}^2) \delta \Phi_\text{cmp}^\dag$
that may produce flux of relativistic particles from the I-ball.
Its effects are discussed in the following as $\delta \Phi_\text{flc}$.

Then, let us focus on remaining modes defined as $\delta \Phi_\text{flc}$,
which might break the non-relativistic condition.
The equation of motion for $\delta \Phi_\text{flc}$ is given by
\begin{align}
\label{eq:fluc}
	\left( \Box + m^2 - \frac{3 \lambda}{2} |\Phi_Q|^2 \right) \delta \Phi_\text{flc} =
	\frac{3 \lambda}{4} \left( \Phi_Q^2 + {\Phi_Q^\dag}^2 \right) \delta \Phi_\text{flc}^\dag.
\end{align}
Since the dispersion relation of $\delta \Phi_\text{flc}^{(\dag)}$ oscillates with time due to the
right hand side of Eq.~\eqref{eq:fluc},
$\delta \Phi_\text{flc}^{(\dag)}$ might grow exponentially due to the parametric resonance.
Roughly speaking, this term may contain two classes of processes: 
number conserving processes ($2$ to $2$)
as the first band
and number violating processes ($n + m$ to $n$) as the higher band.

In the case of gravity mediation type potential, $V_B$ contains infinite
series of terms as
\begin{align}
	V_{\rm B}=\frac{-m^2K}{6}\frac{\Phi^4+{\Phi^\dagger}^4}{|\Phi|^2}
	+\sum_{n\geq3}c_nm^2K\frac{\Phi^{2n}+{\Phi^\dagger}^{2n}}{|\Phi|^{2n-2}},
\end{align}
with $c_n$ being a numerical constant (See Appendix.~\ref{sec:CC}).
The interaction rate that comes from a higher power of $1/|\Phi|^2$ results in
a higher power of $m/|\Phi_Q|$.
Thus, if $m/|\Phi_Q|\ll 1$ holds,
the expansion above is controlled. 
Here and hereafter,
we assume  $m/|\Phi_Q|\ll 1$ 
(in other words $\epsilon^3Q\gg1$)
in the case of gravity mediation type potential.
The interaction terms for $\delta\Phi_{\rm flc}$ can be expressed as the following form
\begin{align}
\label{logint}
	V\supset c\frac{Km^2}{|\Phi|^2}\frac{[\Phi_Q^a \text{ or } {{\Phi_Q^\dagger}^a}]}{|\Phi_Q|^a}
	\frac{(\delta \Phi_{\rm flc}\text{ or }\delta\Phi^\dagger_{\rm flc})^{b}}
	{|\Phi_Q|^{b-4}},
\end{align}
with non negative integer $a$, $b$ and numerical constant c. 
These interaction terms can cause both number conserving and number violating processes
if they are energetically allowed.

The following of this section is devoted to
show\footnote{
A complete analysis on the stability of the I-ball
is quite complicated and beyond the scope of this paper.
Thus, we discuss the stability of the I-ball by an intuitive way.
}
that, for sufficiently long time,
effects of $\delta\Phi_{\rm flc}$ are negligible
and the non-relativistic condition holds.
We estimate the effects from $\delta \Phi_{\rm flc}$ focusing on
particle production processes in two cases:
the charge violating process of number conserving ($n$ to $n$) one and 
number changing ($n+m$ to $n$) one.
%
%
%
%
%
%
%%%%%%%%%%%%%%%%%%%%%%%%%%%%%%%%%%%%%%%%%%%%%%%%%%
\subsection{Number conserving processes}
\label{sec:ncnp}
%%%%%%%%%%%%%%%%%%%%%%%%%%%%%%%%%%%%%%%%%%%%%%%%%%

Let us consider charge violating but number conserving processes.
 In both 
 polynomial potential and gravity mediation type potential,
 the U(1) violating term contains
$V_\text{B} \supset (\Phi^4+{\Phi^\dagger}^4)$, and hence
charge violating but number conserving two to two processes $(++\rightarrow --)$
may happen.

The Q-ball solution with charge Q oscillates with a frequency $\omega(Q)\equiv m-\mu$.
Thus,
the produced particles in 
number conserving processes have energy $\omega(Q)$ per one particle.\footnote{
	If the particles are produced outside the Q-ball, its energy per one particle is larger, $\omega > \omega (Q)$.
	Thus, the below inequality is satisfied much easily.
}
On the other hand, Q-ball solution with charge Q has energy 
$E_{\rm Q-ball}(Q)\simeq\omega(Q)Q$.
Now, suppose that Q-ball solution loses its charge by $\triangle Q$ 
due to number conserving but charge violating processes.
Then, the time-averaged total energy shifts to $\bar{E}'$ 
that is roughly given by
\begin{align}
         \bar{E}'& \simeq \omega(Q-\triangle Q)\cdot[Q-\triangle Q]+\omega (Q)\triangle Q\nonumber
         \\
         &\simeq \omega (Q) Q-\frac{\partial \omega(Q)}{\partial Q}Q\triangle Q \nonumber \\
         & \geq \omega (Q) Q,
\end{align}
if the condition
\begin{align}
\label{cnd:1}
         \frac{\partial \omega(Q)}{\partial Q}<0,
\end{align}
%%
%which is ensured by the shallowness of the potential,
is satisfied.

As explained before,
in the case of gravity mediation potential,
since the charge $Q$ is proportional to $Q \propto m (\Phi_0)\Phi_0^2 R^3 \propto (\Phi_0 / m (\Phi_0))^2$,
the frequency of I/Q-ball $\omega \sim m (\Phi_0)$ decreases as the charge increases.
Therefore, the number conserving but charge violating processes are energetically unfavored.

On the other hand, in the case of polynomial potential,
the  frequency of I/Q-ball decrease as the charge increases
in the flat-top regime 
similar to the case of gravity mediation potential
but the opposite happens in the narrow regime
(see Fig.~\ref{fig:fix}).
Therefore, 
in the narrow regime,
the charge violating but number conserving processes are energetically allowed.
If such a process occurs effectively,
the charge $Q$ drops down
and the shape of the I-ball becomes sharper and sharper,
which makes the non-relativistic condition worse and worse.
Thus, in this regime, 
the collapse of I/Q-balls may be initiated by the charge violating but number conserving processes,
and its time scale can be estimated as $\Gamma\sim \lambda \epsilon^2 \omega$,
which is much longer than the oscillation period
as is consistent with \cite{Fodor:2008es,Hertzberg:2010yz,Amin:2010jq}.
%
%
%
%\subsubsection{With number changing process}
%%%%%%%%%%%%%%%%%%%%%%%%%%%%%%%%%%%%%%%%%%%%%%%%%%
\subsection{Number changing processes}
\label{sec:nchp}
%%%%%%%%%%%%%%%%%%%%%%%%%%%%%%%%%%%%%%%%%%%%%%%%%%

Now, we consider the number changing ($n+m$ to $n$) processes. 
First, let us consider the case of polynomial potential.
Suppose a Q-ball solution with a size $R$,
a wall width $L$,
frequency $\omega$,
and typical amplitude $\Phi_0$.
Using relations Eq.~\eqref{relat}, \eqref{eq:c_q} and $\Phi_0^2\sim\lambda/g$
for flat-top regime, the followings hold 
\begin{align}
         Q&\sim \omega R^3 \Phi_0^2,\\
         \epsilon&\equiv 1/L\omega \sim
         \begin{cases}
         1/Rm~~\text{for narrow regime}\\
         \lambda/\sqrt{g}~~\text{for flat-top regime}
         \end{cases},\\
         \lambda \Phi_0^2&\sim  \epsilon^2 m^2 ~~\text{for both regimes} .
\end{align}
Recall that the parameter $\epsilon\equiv 1/L\omega $
 is small in general [See Eq.~\eqref{eq:small}].
In the case of polynomial potential, 
the dominant charge violating channel is a four to two process with $(++++\rightarrow +-)$
caused by quartic interaction term or sixtic one. 
For produced modes which have momenta compatible with $2\omega$,
the size $R$ and the typical wall width $L$ are rather large.
Thus,
the 
number changing 
particle production processes which have typical length scales $\sim1/\omega$
may not be sensitive to the spatial variations of the amplitude $\partial \ln\Phi/\partial r\sim 1/L\ll \omega$.
Therefore,  for simplicity, we %assume 
approximate that the background condensation is
spatially homogeneous in the following analysis.\footnote{
In some cases, produced particles are expected to escape from the I-ball~\cite{Hertzberg:2010yz,Kawasaki:2013awa}.
In the following, we concentrate on order of magnitude estimation,
and hence the results are correct in this sense.
}
The four to two rate in the homogeneous background condensation
can be calculated by, for example, using a method developed in \cite{Matsumoto:2007rd}.
We can estimate the decay rate of Q-ball charge as
\begin{align}
         \frac{d Q}{dt}\sim 
           -\lambda \epsilon^6\omega Q,
        \end{align}
for quartic interaction and
\begin{align}
        \frac{d Q}{dt}\sim -
           \lambda \epsilon^6\left(\frac{g}{\lambda^2}\right)^2\omega Q,
\end{align}
for sixtic interaction.
Finally one can get typical decay rate by this process as
%%
%\begin{align}
 %        \frac{d\ln Q}{dt}\sim -\lambda\epsilon^6 \omega,
%\end{align}
%%
%%
\begin{align}
\label{eq:decay}
         \Gamma_{I,4\rightarrow 2}\sim 
         \begin{cases}
         \lambda \epsilon^6 \omega~~~~ \text{by quartic interaction}\\
       \lambda \epsilon^6\left(\frac{g}{\lambda^2}\right)^2\omega \lesssim
           \lambda \epsilon^2 \omega
           ~~\text{by sixtic interaction}
           \end{cases},
\end{align}
which is suppressed by factors on $\lambda$ and $\epsilon$ compared to the frequency of I-ball.
This result agrees with oscillon decay rate with a quartic or sixtic interaction estimated in
\cite{Hertzberg:2010yz}.
One can say that smallness of 
the parameter $\epsilon\equiv 1/L\omega $,
which is a consequence of the non-relativistic condition,
ensures the approximate
stability of I-ball.

Let us see the case of gravity mediation type potential.
In this case, interaction terms have the form of \eqref{logint}.
Using dimensional analysis we can estimate typical decay rate as
\begin{align}
	\Gamma_{\rm typ}\sim \frac{\epsilon^{1+4d}}{Q}(\epsilon^3 Q)^{-e}\omega,
\end{align}
with non negative integer $d$ and $e$ depending on the interactions.
The largest rate comes from the reaction $(++++\rightarrow ++)$ with
vertex $\sim Km^2\Phi^6/|\Phi|^2$ with $d=e=0$.
With $\epsilon^3Q \gg 1$, the rate above is highly suppressed.

%%%%%%%%%%%%%%%%%%%%%%%%%%%%%%%%%%%%%%%%%%%%%%%%%%
\section{Discussion and Conclusions}
\label{sec:conc}

In a non-relativistic configuration, a classical real scalar field theory
can be embedded into a complex one with a conserved U(1) charge.
From this fact, we have shown that
an I-ball can be understood as a
projection of a Q-ball, 
and that if the non-relativistic condition holds, the stability of I-ball is guaranteed by
that of Q-ball.
We have discussed the robustness of the non-relativistic condition
and confirmed the
long-lived nature of I-ball.
Interestingly, we have seen that
the main processes which determine a lifetime of the I/Q-ball may
depend on the property of its frequency as a function of charge, $\omega (Q)$:
number conserving processes are unfavored/favored for
$\der \omega / \der Q \lessgtr 0$.

As one can guess, the separation of time scale plays a crucial role in the formation of I-ball.
If one prepares inhomogeneous non-relativistic real scalar fields with a potential nearly but shallower than quadratic,
they first fragment into I-balls if the non-relativistic condition
is maintained for sufficiently long time. 
Then, they gradually reduce their number by producing relativistic particles,
and the system finally relaxes into thermal equilibrium.

In more realistic situations,
the  background space-time expands,
and a real scalar field may interact with light particles which participate in the background thermal plasma.
On the one hand,
since the momenta of scalar fields decrease due to the expansion of the Universe,
the non-relativistic condition is expected to be satisfied easily.
On the other hand,
the interaction with the background plasma might make the scalar field energetic,
and eventually the I-ball may evaporate~\cite{Mukaida:2012qn}.
It is also interesting to pursue, with what initial conditions
and external force that deviates the system from thermal equilibrium 
({\it e.g.}, the cosmic expansion),
the scalar field first fragments into the I-ball
not into thermal equilibrium particles~\cite{Berges:2012us,Nowak:2013juc,Berges:2014xea,Davidson:2014hfa}.
We will return to these issues elsewhere.

%%%%%%%%%%%%%%%%%%%%%%%%%%%%%%%%%%%%%%%%%%%%%%%%%%

%%%%%%%%%%%%%%%%%%%%%%%%%%%%%%%%%%%%%%%%%%%%%%%%%%
\section*{Acknowledgment}
%%%%%%%%%%%%%%%%%%%%%%%%%%%%%%%%%%%%%%%%%%%%%%%%%%
We would like to thank Masaki Yamada for useful discussion.
This work is supported by Grant-in-Aid for Scientific
research from the Ministry of Education, Science, Sports, and Culture
(MEXT), Japan.
The work of K.M. and M.T. is supported in part by JSPS Research Fellowships
for Young Scientists.
The work of M.T. is supported by the program
for Leading Graduate Schools, MEXT, Japan.

%--------------------------------------------------------------------------%
\appendix

%%%%%%%%%%%%%%%%%%%%%%%%%%%%%%%%%%%%%%%%%%%%%%%%%%
\section{Embedding a real scalar field theory into complex one}
\label{sec:CC}

We consider a real scalar $\phi$ field theory with the potential $V(2\phi^2)$.
The equation of motion is the following
\begin{align}
         \Box \phi+4V'(2\phi^2)\phi=0.
\end{align}
Now, we introduce a complex field $\Phi$ which satisfies $\Re(\Phi)=\phi$.
The following holds
\begin{align}
         4V'(2\phi^2)\phi=2V'\left(|\Phi|^2+\frac{\Phi^2+{\Phi^{\dagger}}^2}{2}\right)
         \cdot(\Phi+\Phi^\dagger).
\end{align}
By expanding the potential around $|\Phi|^2$, one can obtain the following form
\begin{align}
         4V'(2\phi^2)\phi&=2V'
         \left(|\Phi|^2+\frac{\Phi^2+{\Phi^{\dagger}}^2}{2}
         \right)(\Phi+\Phi^\dagger)
         \nonumber \\
         &=\sum_{n}k_n(|\Phi|^2)\left(\frac{\Phi^2+{\Phi^{\dagger}}^2}{2}\right)^n
         (\Phi+\Phi^\dagger)\nonumber \\
         &=\sum_{n}g_n(|\Phi|^2) \left[\Phi^{2n+1}+{\Phi^{\dagger}}^{2n+1}\right],
\end{align}
with some functions $k_n$ and $g_n$.
Suppose $\Phi=re^{i\theta}$, we have
\begin{align}
         2rV'\left(r^2(1+\cos 2\theta)\right)
         \cos\theta=\sum_ng_n(r^2)\cos((2n+1)\theta)r^{2n+1}.
\end{align}
From this expression, we can calculate the function $g_n(x)$ 
\begin{align}
         g_n(r^2)=\frac{2}{\pi}\int_{-\pi}^{\pi}\frac{
         V'\left(r^2(1+\cos 2\theta)\right)
         \cdot\cos\theta\cdot\cos((2n+1)\theta)
         }{r^{2n}}d\theta.
\end{align}
On the other hand,  the equation of motion can be rewritten as
\begin{align}
         \Re\left[\Box \Phi+\sum_{n}g_n(|\Phi|^2)\left(\Phi^{2n+1}+{\Phi^{\dagger}}^{2n+1}\right)\right]
        +\Re\left[\sum_{n}f_n(|\Phi|^2)\left(\Phi^{2n+1}-{\Phi^{\dagger}}^{2n+1}\right)\right]=0,
\end{align}
with arbitrary functions $f_n(x)$
 because the last term vanishes. We can consider a complex scalar equation of motion
\begin{align}
         \Box \Phi+\sum_{n}\left[g_n(|\Phi|^2)+f_n(|\Phi|^2)\right]\Phi^{2n+1}+
         \sum_{n}\left[g_n(|\Phi|^2)-f_n(|\Phi|^2)\right]{\Phi^\dagger}^{2n+1}=0.
\end{align}
This equation of motion can be derived from the potential $G(\Phi,\Phi^\dagger)$ which satisfies
\begin{align}
         G(\Phi,\Phi^\dagger)&=\sum_n G_n(|\Phi|^2)\left[\Phi^{2n}+{\Phi^\dagger}^{2n}\right],\\
         g_n(x)+f_n(x)&=G'_n(x),\\
         g_n(x)-f_n(x)&=xG'_{n+1}(x)+2(n+1)G_{n+1}(x),\\
         G_1(x)&=0.
\end{align}
With condition $G_1(x)=0$, we can compute $f_n$ and $G_n$. This is because
$G_{n+1}$ can be determined by
\begin{align}
         2(n+1)G_{n+1}(x)+xG'_{n+1}(x)=2g_n(x)-G'_n(x).
\end{align}
%%
%where we assume $G_{n+1}(x)$ can have a solution for arbitral left hand side.
%%
Now, we have the charge conserving potential $V_{\rm U(1)}$ and violating term $V_{\rm B}$ as
\begin{align}
         V_{\rm U(1)}(|\Phi|^2)&\equiv G_0(|\Phi|^2),\\
         V_{\rm B}(\Phi,\Phi^\dagger)&\equiv \sum_{n\geq2}G_n(|\Phi|^2) 
         \left[\Phi^{2n}+{\Phi^\dagger}^{2n}\right].
\end{align}
For example,
in the case of polynomial  potential $V(x)=Ax^{m+1}/(m+1)$ with constant A,
 $G_n(x)$ can be obtained as
\begin{align}
         G_n(x)=2^{2-m}A\frac{1+(-1)^n}{2}\frac{{}_{2m+1} \mathrm{C} _{m-n+1}
         }{m+n+1}x^{m+1-n}\text{    for $n\leq m+1$  otherwise zero.}
\end{align}
%%

%{\bf (2) log : $V(x)=-m^2xK\ln\frac{x}{\Lambda^2}$ with constant $m,K$}
%%%%%%%%%%%%%%%%%%%%%%%%%%%%%%%%%%%%%%%%%%%%%%%%%%

%%%%%%%%%%%%%%%%%%%%%%%%%%%%%%%%%%%%%%%%%%%%%%%%%%

%%%%%%%%%%%%%%%%%%%%%%%%%%%%%%%%%%%%%%%%%%%%%%%%%%

\end{document}